\documentclass[letterpaper, 11pt, compsoc]{IEEEtran}
\usepackage[margin=1.1in]{geometry}
\usepackage{graphicx}
\graphicspath{{Figures_Cryptosystem/}}
\usepackage{amssymb,amsmath}
\usepackage{amsthm}
\usepackage[ruled,vlined,linesnumbered]{algorithm2e}
\usepackage{multirow}
\usepackage{subcaption}
\usepackage{color}
\usepackage{rotating}
\usepackage{balance}
\usepackage{fancyhdr}

\begin{document}

\title{Personal Internet of Things (PIoT): \\What is it Exactly?}

\author{
	\begin{flushleft}
		\begin{tabular}{llll}
			\\
\textbf{Biswa P. S. Sahoo}  & & & \textbf{Saraju P. Mohanty} \\
Samsung R\&D Institute Bangalore & & &  University of North Texas \\ \\ \\ 
\textbf{Deepak Puthal} & & & \textbf{Prashant Pillai} \\
Newcastle University & & & University of Wolverhampton \\
		\end{tabular}
	\end{flushleft}
}


\maketitle


\begin{abstract}
The use of Internet of Things (IoT) devices in homes and the immediate proximity of an individual communicates to create Personal IoT (PIoT) networks. The exploratory study of PIoT is in its infancy, which will explore the expansion of new use cases, service requirements, and the proliferation of PIoT devices. This article provides a big picture of PIoT architecture, vision, and future research scope.
\end{abstract}


\section*{Towards a Hyper-Connected World}
We are moving towards a hyper-connected world through the evolution of Internet-of-Things (IoT).~\cite{mohanty2016everything}. The widely used "Personal Network" enables the simplification of the network functionalities, for example, user session redirection between devices such as phone, Personal Digital Assistants (PDA), and laptop. However, in the present day, the most crucial use case for IoT devices in the consumer segment is becoming consumer internet and media devices such as smartphones, where the number of IoT devices is expected to thrive to more than eight billion by 2030~\cite{guo2021enabling}.

The existing mobile IoT standards, such as Narrow-Band IoT (NB-IoT) and enhanced Machine-Type Communication (eMTC) to support many IoT verticals~\cite{sahoo2018enabling}. However, these mobile IoT standards, so far, has not looked at the proliferation of consumer IoT devices that can be categorized into two broad areas \cite{piotstudy}: i) IoT in Home (such as door sensors, cameras, smart TVs, and refrigerator) and ii) IoT on Person (such as wearable, smartphone, car, and handheld devices).


\section*{Personal Internet of Things}
The above mentioned two categories of IoT devices (i.e., IoT in Home and IoT on Person) that communicate in the order of a meter or a couple of meters between themselves and with an external network via a local gateway are collectively called the Personal IoT (PIoT) networks. For example, a user with a smartphone or any other IoT device sitting in the car gets connected to the car creates a PIoT network. Thus, we define the PIoT network as ``\emph{A group of connected devices focused mainly in homes and the immediate proximity of an individual}". 

The recent technological advancement in antenna technology, massive MIMO, communication on higher spectrum bands provides high datarate wireless broadband, long and short-range communications. This technological progression has led to various applications relating to consumer goods. Nonetheless, when it comes to extreme short-range communications, as in the case of the PIoT network, the current mobile IoT standards or 5G standards, in a broad sense, do not support extreme short-range communication capabilities. Therefore, the PIoT network demands a framework that enables the seamless integration of PIoT into the 5G ecosystem. 


\begin{figure}[t]
	\centering
	\includegraphics[width=\columnwidth]{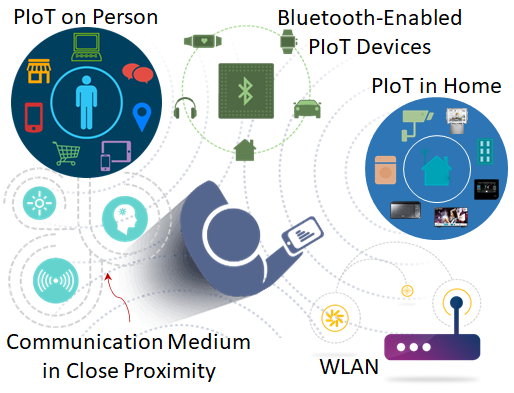}
	\caption{An abstract view of PIoT networks.}
	\label{fig:piotmain}
\end{figure}

\section*{Creating PIoT Networks}
\begin{figure}[t]
	\centering
	\includegraphics[width=\columnwidth]{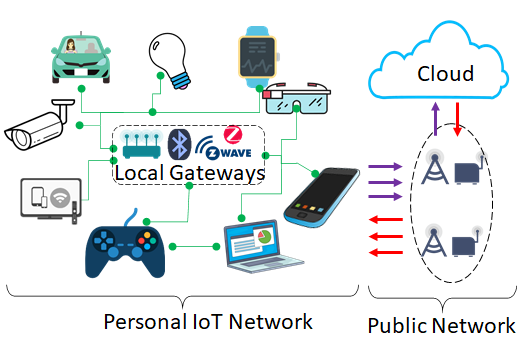}
	\caption{Personal IoT networks in a vicinity}
	\label{fig:piotnetwork}
\end{figure}

IoT devices, mainly homes or around an individual's vicinity such as home, car, and office, create personal networks, as shown in Fig.~\ref{fig:piotmain}. These personal devices use different non-3GPP-based technologies (such as Bluetooth, ZigBee, Z-Wave, Infrared, and NFC) as a local gateway, as shown in Fig.~\ref{fig:piotnetwork}. In PIoT networks, each of the IoT devices or application might require a different types of connectivity methods. For example, garage door openers use Z-Wave or Zigbee and require a home base control unit, whereas personal voice assistance might use WLAN technologies; and needs constant interaction to a smartphone using cloud-based control. Thus, the requirements for creating a PIoT network vary considerably from device to device and day-to-day operation. Furthermore, the communication complexity and cost-effectiveness are also challenging. Consequently, the seamless integration of the PIoT devices communicating over different non-3GPP-based technologies into the 3GPP-based network is crucial.

In PIoT, we broadly classify two potential issues: i) device to device within the PIoT network and ii) interaction between PIoT devices and 3GPP network. Firstly, as the devices are neither fully managed by the PIoT network nor by the 3GPP network, setting up the PIoT network is difficult and costly at times. Secondly, the IoT device connectivity is so much fragmented, as shown in Fig.~\ref{fig:piotnetwork}. To summarise, this fragmentation confuses the user and makes it difficult to take the mass adaptability of consumer PIoT devices and provide seamless connectivity. Current age technology like Blockchain, Artificial Intelligence,
Edge Computing, Data Analysis will strengthen
sustainable and secure PIoT deployment.

\section*{Communication Aspects of Personal IoT}
The current 5G standards support both short and long-range communication; however, they do not support very short-range communications and consumer-based private networks (e.g., PIoT networks) where a user can, in conjunction with an operator (PLMN) can easily manage consumer IoT devices \cite{piotstudy}. Thus, the support for PIoT in the 5G standard is essential to proliferate consumer IoT devices and PIoT networks. The future study of 3GPP must focus on PIoT to support the service and functional requirements, identify the characteristics of PIoT networks, and simplify the network architecture to enable the 3GPP network to support PIoT functionalities seamlessly. Besides, the 5GS also must study the gap analysis between the identified requirements and what is already defined by existing 3GPP requirements.


\section*{Cybersecurity Aspects of Personal IoT}
Possible security threats in PIoT are comparatively lower than the IoT due to its limited exposure to public networks. Unlike IoT, we can categorise the PIoT threats in the two parts, i.e., device level, communication level.

Device capture, device tampering, and device outage are prevalent attacks in device-level threats. On the one hand, PIoT has limited chances to be exposed to device-level attacks; however, it cannot be ignored altogether; on the other hand, PIoT is prone to communication-level attacks. Generally, PIoT devices communicate to the Cloud data center through cellular network gateways. Usually, these gateways are open to public networks and may lead to network attacks, such as selective forwarding, blackhole, wormhole attack \cite{puthal2019fog}. Existing security solutions may be taken as a base to secure the PIoT \cite{puthal2019fog}; however, it demands new security solutions with application specifications.


\section*{Energy Aspects of Personal IoT}
Energy-aware is a crucial design goal in a wide range of IoT devices such as wearables and sensors. However, the current IoT standards do not support very short-range low energy consumption capabilities. Thus, enhancements in energy consumption optimization for PIoT devices become essential. One way to go with this is to optimize the data traffic that can provide significant energy saving, e.g., small data transmission. Besides, a gap analysis requires between the identified requirements for PIoT and what is already defined for IoT requirements.

\section*{Conclusion}

The PIoT comprises a set of IoT devices that communicate between themselves and with a local gateway. Although communication among IoT devices is not entirely new, but creating a PIoT networks requires lots of nonautomated configurations, to which a user might find difficult. Thus, the current state of IoT devices requires a regulatory framework in 5GS for PIoT.


\bibliographystyle{ieeetran}
\bibliography{PIoT_CEM}
\par
\noindent \textbf{Biswa P. S. Sahoo} is a Senior Researcher at the Samsung R\&D Institute Bangalore (SRIB), India. Contact him at: biswa.p@samsung.com. \\\\
\textbf{Deepak Puthal} is an Assistant Professor at the Newcastle University, UK. Contact him at: deepak.puthal@newcastle.ac.uk. \\\\
\textbf{Saraju P. Mohanty} is a Professor at the University of North Texas, Denton, USA. Contact him at: smohanty@ieee.org. \\\\
\textbf{Prashant Pillai} is a professor at the University of Wolverhampton, UK. Contact him at: p.pillai@wlv.ac.uk.

\end{document}